\documentclass[aip,apl,a4,reprint,twoside,floatfix,superscriptaddress]{revtex4-1} %as an APL paper

\usepackage{graphicx}
\usepackage{amsmath}
\usepackage{amssymb}
\usepackage{amsfonts}
\usepackage{dcolumn}
\usepackage{epsfig}
\usepackage{subfigure}
\usepackage{bm}

\begin{document}

\title{\bf Multiferroic coupling in nanoscale BiFeO$_3$}

\author {Sudipta Goswami} \affiliation {Nanostructured Materials Division, Central Glass and Ceramic Research Institute, CSIR, Kolkata 700032, India}
\author {Dipten Bhattacharya}
\email{dipten@cgcri.res.in} \affiliation {Nanostructured Materials Division, Central Glass and Ceramic Research Institute, CSIR, Kolkata 700032, India}
\author {P. Choudhury} \affiliation {Nanostructured Materials Division, Central Glass and Ceramic Research Institute, CSIR, Kolkata 700032, India}
\author {B. Ouladdiaf} \affiliation {Science Division, Institut Laue-Langevin, Bo$\hat{i}$te Postale 156, 38042 Grenoble Cedex 9, France}
\author {T. Chatterji} \affiliation {Science Division, Institut Laue-Langevin, Bo$\hat{i}$te Postale 156, 38042 Grenoble Cedex 9, France}

\date{\today}

\begin{abstract}
Using the results of x-ray and neutron diffraction experiments, we show that the ferroelectric polarization, in $\sim$22 nm particles of BiFeO$_3$, exhibits a jump by $\sim$30\% around the magnetic transition point $T_N$ ($\sim$635 K) and a suppression by $\sim$7\% under 5T magnetic field at room temperature ($\ll$$T_N$). These results confirm presence of strong multiferroic coupling even in nanoscale BiFeO$_3$ and thus could prove to be quite useful for applications based on nanosized devices of BiFeO$_3$. 
\end{abstract}

\pacs{75.80.+q, 75.75.+a, 77.80.-e}
\maketitle

The hope in the multiferroicity of BiFeO$_3$ has been rekindled by the observation of spin-flop under electric field in a single crystal.\cite{Lebeugle} This result puts to rest the debate\cite{Ederer} whether BiFeO$_3$, which is otherwise so attractive because of its multiferroicity at room temperature, depicts a weak or strong coupling between ferroelectric and magnetic order parameters. While efforts are on to improve the coupling even further via designed thin film structures,\cite{Catalan} questions naturally arise about how nanosize can influence the coupling. It has been shown\cite{Fong} in recent times that long-range ferroelectric order persists in just 3 unit-cell thick ($\sim$1.2 nm) films of PbTiO$_3$. For nanosized BiFeO$_3$ too, we\cite{Goswami} have already shown that substantial polarization is retained in particles as fine as $\sim$20 nm. And, in spite of some controversies,\cite{Bea} it is now generally accepted that intrinsic ferromagnetism emerges in particles finer than $\sim$62 nm due to suppression of the spin spiral.\cite{Park,Mazumder,Wang} However, the question of multiferroic coupling in nanoscale BiFeO$_3$ has not been addressed adequately.
 
In this paper, we show from x-ray and neutron diffraction experiments that not only sizable ferroelectric polarization is retained in nanosized ($\sim$22 nm) particles of BiFeO$_3$, the coupling between magnetic and electric order parameters too, is quite strong in them.

The nanoparticles of BiFeO$_3$ have been synthesized by sonochemical route where coprecipitation takes place within a suitable medium from aqueous solution of mixed metal nitrates in presence of ultrasonic vibration. The details of the powder preparation and their characterization have been given elsewhere.\cite{Goswami} In Fig.1a,b, we show a representative transmission electron micrograph (TEM) and a high resolution TEM (HRTEM) photograph. The HRTEM photograph establishes the single-crystalline nature of the particles. The lattice fringes in Fig.1b correspond to the crystallographic (012) plane. Either (012) or (110) planes were found to be oriented with respect to the beam direction in almost all the particles examined by HRTEM. The magnetic transition point $T_N$ has been noted from dc magnetization measurements across 300-900 K. The magnetic transition is associated with a characteristic endothermic peak in calorimetry. In particles of average size $\sim$22 nm, $T_N$ reduces down to $\sim$635 K from $\sim$653 K in bulk.\cite{Goswami} We have carried out high resolution powder x-ray diffraction experiment using the laboratory x-ray diffractometer (with CuK$\alpha$ source and monochromator) across $\sim$300-700 K (Fig. 2). In addition, we have also carried out powder neutron diffraction under zero and 5T magnetic field at room temperature (Fig. 3) at the D20 diffractometer of Institut Laue-Langevin (ILL). 

\begin{figure}[!h]
  \begin{center}
    \includegraphics[scale=0.35]{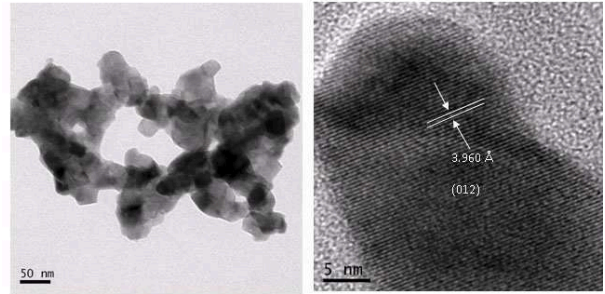} 
    \end{center}
  \caption{(a) TEM photograph of the nanosized particles of BiFeO$_3$ and (b) HRTEM photograph showing the lattice fringes for (012) plane. }
\end{figure}
 
\begin{figure}[!h]
\centering
\includegraphics[scale=.28]{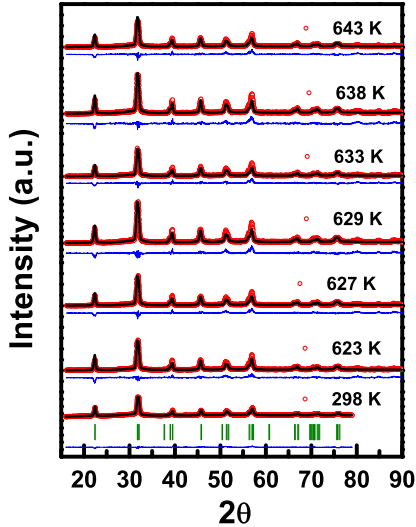}
\caption {(color online) The experimental and FullProf refined x-ray diffraction patterns for nanoscale BiFeO$_3$ at different temperatures; red colored open circles represent the experimental data points, black solid lines represent the calculated data, blue solid line represent the difference between observed and calculated data, and green lines represent the Bragg lines. The fit statistics and other relevant details are available in the supplementary information.} 
\end{figure}

\begin{figure}[!h]
\centering
\includegraphics[scale=.30]{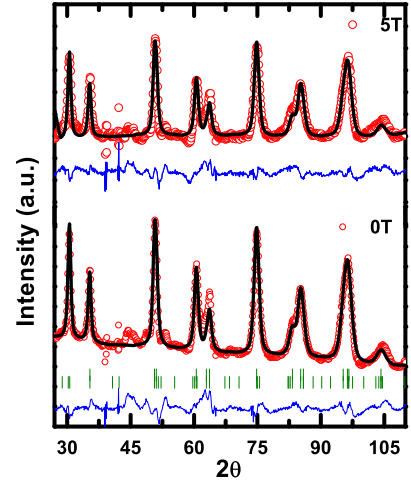}
\caption {(color online) The experimental and FullProf refined neutron diffraction patterns ($\lambda$ = 2.41 \AA) for nanoscale BiFeO$_3$ at room temperature and under zero and 5T magnetic field; red colored open circles represent the experimental data points, black solid lines represent the calculated data, blue solid line represent the difference between observed and calculated data, and green lines represent the Bragg lines for both magnetic and crystallographic lattices. The fit statistics and other relevant details are available in the supplementary information.} 
\end{figure}

In Fig. 2, we show the experimental and FullProf refined x-ray diffraction patterns at several temperatures across $T_N$. A small amount of Bi$_2$Fe$_4$O$_9$ impurity ($\sim$5$\%$) is found to be present. The refinement has been done by taking into consideration the microstructual effects - finite crystallite size and strain - as well. Keeping the instrumental Gaussian parameters ($U$, $V$, $W$) for peak broadening constrained at the values obtained for standard crystal (LaB$_6$), other parameters such as Gaussian and Lorentzian parameters for peak broadening due to crystallite size and strain, atom positions, asymmetry parameters etc have been relaxed and appropriate isotropic and anisotropic size models have been invoked. The detail results of the refinement - lattice constants, atom positions, Bi-O and Fe-O bond lenghts, average $<$Fe-O-Fe$>$, $<$Bi-O-Bi$>$ bond angles together with fit statistics such as R-factors, $\chi$$^2$ etc - have been given in the supplementary information.\cite{supplementary} The space group turns out to be R3c in the nanoscale system at room temperature and across $T_N$ similar to what has been observed in the bulk. The bulk BiFeO$_3$, of course, undergoes a phase transition around the ferroelectric Curie point $\sim$1103 K from rhombohedral R3c to orthorhombic Pbnm.\cite{Arnold,Knight} Using the atom position coordinates, the off-center displacements of Bi$^{3+}$ ($\delta_{Bi}$) and Fe$^{3+}$ ($\delta_{Fe}$) ions within the Bi-O and Fe-O cages of a unit cell have been determined. Bi$^{3+}$ is found to occupy the tetrahedral site while Fe$^{3+}$ occupies the octahedral site. The net unit-cell polarization has been estimated from the off-center displacement ($\delta$) using the procedure laid down in Ref. 13. The direction of the polarization is found to be canted away from the [111]$_{rh}$$\parallel$[001]$_{hex}$ axis by $\sim$27$^o$ toward the ab-plane of hexagonal setting. The reason for the canting could be the large strain in the particles ($\sim$0.1\%).\cite{Goswami,Jang} However, the direction switches toward [111]$_{rh}$$\parallel$[001]$_{hex}$ with the rise in temperature from below to $T_N$. But, most interestingly, the polarization ($P$) is found to exhibit a clear anomaly around $T_N$ (Fig. 4e). $P$ rises by $\sim$30\% across $T_N$. The lattice parameters $a$, $c$ and the lattice volume $V$ too, show distinct anomaly at $T_N$ (Figs. 4b,c). We note the anomalous changes in the lattice parameters $a$ and $c$ across $T_N$ which could arise from the magnetoelastic effect. The volume contraction ($\sim$0.4\%) at $T_N$ establishes the first order nature of the transition (Fig. 4c). It corroborates the calorimetry data (Fig. 4a). For even finer particles, of course, latent heat is found to be zero indicating rounding off due to enhanced surface effects.   

\begin{figure}[!h]
\centering
\includegraphics[scale=.28]{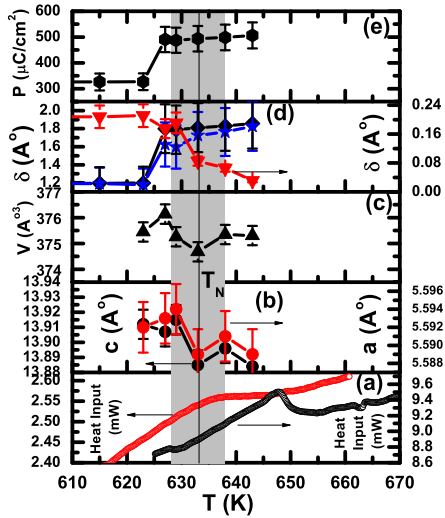}
\caption {(color online) (a) DSC traces for the bulk (black) and nanosized particles (red) of BiFeO$_3$ around T$_N$; the peak in the nanoscale sample is smeared pointing out a broader transition zone which is marked in the plot; (b) variation of lattice parameters a (right axis) and c (left axis) with temperature; (c) variation of lattice volume with temperature; volume contraction takes place around T$_N$; (d) variation of $\delta_{Bi}$ (diamond) and $\delta_{Fe}$ (down traingle) as well as the net $\delta$ (star) in a unit cell; (e) variation of ferroelectric polarization with temperature. } 
\end{figure}

In order to verify the coupling, we have carried out powder neutron diffraction under zero and 5T magnetic field at room temperature. In Fig. 3, we show the experimental and refined neutron diffraction patterns for the $\sim$22 nm particles of BiFeO$_3$. The propagation vector is found to be (0,0,0) indicating commensurate magnetic lattice. The details of the refinement have been given in the supplementary information.\cite{supplementary} The magnetization per Fe$^{3+}$ ion turns out to be $\sim$3.22 $\mu$$_B$. This result corroborates the results obtained by others.\cite{Pandey} We, of course, emphasize here more on the $\textit{change}$ in net $\delta$ under 5T magnetic field; $\delta$ is found to have decreased by $\sim$0.06 \AA which amounts to a suppression in polarization by $\sim$7\%. No switching in the direction of polarization under magnetic field could be observed in this case. Of course, since the experiments have been carried out on powdered sample, it is not possible to ascertain the direction of the magnetic field with respect to the crystallographic axes. It is important to mention here that the structural details determined both from the x-ray and neutron diffraction data at 300 K and under zero magnetic field corroborate each other indicating consistency in the data as well as their refinement. 

The sharp change ($\sim$30\%) in the polarization around $T_N$ as well as its suppression ($\sim$7\%) under 5T magnetic field at a temperature well below $T_N$ point out that the multiferroic coupling is quite strong even in nanoscale ($\sim$22 nm) BiFeO$_3$. It is also noteworthy that the onset of magnetic order influences the polarization more than the application of magnetic field at well below $T_N$. Although, the reason behind such observation is not quite clear at this moment, it apparently points out that the domain switching and consequent striction mediated coupling between ferroelectric and magnetic domains - as observed in hexagonal manganites\cite{Fiebig} - is, perhaps, not the dominant mechanism here. It has been shown already that the antiferromagnetic spin structure with spiral modulation and small moment $\sim$0.02 $\mu$$_B$, as observed in bulk or thin film system,\cite{Eerenstein} could be suppressed by reducing the size below $\sim$62 nm,\cite{Park} increasing the strain, doping,\cite{Zalesski} and applying high magnetic field.\cite{Ruette} The suppression of spin spiral in nanoscale BiFeO$_3$ yields large magnetization\cite{Mazumder} ($\sim$0.4 $\mu_B$/Fe) which, in turn, gives rise to an enhanced nonferroelectric rotation\cite{Ederer} of oxygen octahedra about the [111] axis via Dzyaloshinskii-Moriya interaction. Since it has already been shown\cite{Ederer} that the polar and rotational distortions are coupled, it could be the origin of the observed multiferroic coupling in nanoscale BiFeO$_3$. Of course, it remains to be explored whether the coupling observed in this work is retained in a nanoscale device of BiFeO$_3$ if ferroelectric polarization is measured directly by electrical P-E loop tracing under zero and a finite magnetic field. This will be attempted in a future work.

In conclusion, we show that the ferroelectric polarization in nanoscale ($\sim$22 nm) BiFeO$_3$ exhibits an anomalous change at $T_N$ and a suppression under a magnetic field at below $T_N$. This result confirms that multiferroic coupling is quite strong even in nanoscale which could prove to be useful for developing nanoscale devices of BiFeO$_3$. 

This work has been supported by the CSIR networked research program $"$Nanostructured Advanced Materials$"$ (NWP-051).

\begin{table*} [!h]
\caption{Structural details and fit statistics from FullProf refinement of x-ray diffraction data.}

\begin{tabular}{p{0.3in}p{0.8in}p{0.4in}p{0.5in}p{0.6in}p{0.6in}p{0.6in}p{0.4in}p{0.5in}p{0.5in}p{0.6in}p{0.3in}p{0.3in}p{0.3in}} \hline \hline \\

T \newline \centering (K) & Lattice \newline Parameters \newline \centering (\AA) & Ions & Wycoff \newline Positions & x & y & z & Bonds & Length \newline \centering (\AA) & Bonds & Angle \newline \centering ($^o$)& $R_p$ & $R_{wp}$ & $\chi^2$ \\ \hline 
\\
298 & a = 5.580(9) \newline c = 13.867(7) & Bi \newline Fe \newline O & 6a \newline 6a \newline 18b & 0.0 \newline 0.0 \newline 0.4599(3) & 0.0 \newline 0.0 \newline 0.0139(2) & 0.0 \newline 0.2235(2) \newline -0.0409(5) & Bi-O \newline Bi-O \newline Fe-O \newline Fe-O & 2.591(2) \newline 2.344(3) \newline 1.941(3) \newline 2.070(4) & Fe-O-Fe \newline O-Bi-O & 161.08(2) \newline 70.68(1) & 17.5 & 22.4 & 2.33 \\  
\\
623 & a = 5.592(5) \newline c = 13.912(7) & Bi \newline Fe \newline O & 6a \newline 6a \newline 18b & 0.0 \newline 0.0 \newline 0.4607(6) & 0.0 \newline 0.0 \newline 0.0122(4) & 0.0 \newline 0.2263(4) \newline -0.0388(2) & Bi-O \newline Bi-O \newline Fe-O \newline Fe-O & 2.376(2) \newline 2.599(5) \newline 1.941(2) \newline 2.087(3) & Fe-O-Fe \newline O-Bi-O & 161.32(2) \newline 70.09(4) & 16.5 & 21.3 & 2.41 \\ 
 \\ 
627 & a = 5.593(4) \newline c = 13.907(8) & Bi \newline Fe \newline O & 6a \newline 6a \newline 18b & 0.0 \newline 0.0 \newline 0.4656(3) & 0.0 \newline 0.0 \newline -0.0304(5) & 0.0 \newline 0.2254(5) \newline -0.0372(3) & Bi-O \newline Fe-O \newline Fe-O & 2.554(3) \newline 2.022(1) \newline 2.078(4) & Fe-O-Fe \newline O-Bi-O & 172.93(4) \newline 74.64(2) & 16.5 & 21.1 & 2.26 \\ 
 \\ 
629 & a = 5.594(5) \newline c = 13.915(7) & Bi \newline Fe \newline O & 6a \newline 6a \newline 18b & 0.0 \newline 0.0 \newline 0.4619(4) & 0.0 \newline 0.0 \newline -0.0212(6) & 0.0 \newline 0.2275(5) \newline -0.0379(5) & Bi-O \newline Fe-O \newline Fe-O & 2.494(5) \newline 1.791(3) \newline 2.201(3) & Fe-O-Fe \newline O-Bi-O & 169.89(6) \newline 74.26(4) & 16.6 & 21.2 & 2.21 \\ 
 \\
633 & a = 5.589(7) \newline c = 13.885(6) & Bi \newline Fe \newline O & 6a \newline 6a \newline 18b & 0.0 \newline 0.0 \newline 0.4547(4) & 0.0 \newline 0.0 \newline 0.0454(3) & 0.0 \newline 0.2199(2) \newline -0.0362(2) & Bi-O \newline Fe-O \newline Fe-O & 2.582(2) \newline 1.764(1) \newline 2.219(3) & Fe-O-Fe \newline O-Bi-O & 171.86(5) \newline 76.13(4) & 16.5 & 24.2 & 2.58 \\ 
 \\  
638 & a = 5.591(3) \newline c = 13.896(7) & Bi \newline Fe \newline O & 6a \newline 6a \newline 18b & 0.0 \newline 0.0 \newline 0.4705(5) & 0.0 \newline 0.0 \newline -0.0375(7) & 0.0 \newline 0.2197(3) \newline -0.0350(3) & Bi-O \newline Fe-O \newline Fe-O & 2.565(4) \newline 1.803(2) \newline 2.174(5) & Fe-O-Fe \newline O-Bi-O & 175.34(4) \newline 74.69(2) & 16.0 & 23.5 & 2.68 \\ 
 \\
643 & a = 5.589(5) \newline c = 13.884(8) & Bi \newline Fe \newline O & 6a \newline 6a \newline 18b & 0.0 \newline 0.0 \newline 0.4691(7) & 0.0 \newline 0.0 \newline -0.0326(9) & 0.0 \newline 0.2195(3) \newline -0.0328(5) & Bi-O \newline Fe-O \newline Fe-O & 2.571(3) \newline 1.843(5) \newline 2.136(5) & Fe-O-Fe \newline O-Bi-O & 173.90(6) \newline 73.38(5) & 15.8 & 23.3 & 2.59 \\ 
 \\  
\hline \hline

\end{tabular}
\end{table*}

\begin{table*} [!h]
\caption{Structural details and fit statistics from FullProf refinement of neutron diffraction data.}

\begin{tabular}{p{0.2in}p{0.8in}p{0.4in}p{0.6in}p{0.6in}p{0.6in}p{0.6in}p{0.4in}p{0.5in}p{0.5in}p{0.6in}p{0.3in}p{0.3in}p{0.3in}} \hline \hline \\

Field (T) & Lattice \newline Parameters \newline \centering (\AA) & \centering Ions & Wycoff \newline Positions & x & y & z & Bonds & Length \newline \centering (\AA) & Bonds & Angle \newline \centering ($^o$)& $R_p$ & $R_{wp}$ & $\chi^2$ \\ \hline 
\\
0 & a = 5.607(4) \newline c = 13.901(5) & Bi \newline Fe \newline O & 6a \newline 6a \newline 18b & 0.0 \newline 0.0 \newline 0.4524(7) & 0.0 \newline 0.0 \newline 0.0183(5) & 0.0 \newline 0.2213(5) \newline 0.9412(4) & Bi-O \newline Bi-O \newline Fe-O \newline Fe-O & 2.171(3) \newline 2.615(5) \newline 1.891(5) \newline 2.216(3) & Fe-O-Fe \newline O-Bi-O & 151.55(5) \newline 77.14(4) & 18.9 & 16.8 & 30.3 \\  
\\
5 & a = 5.605(6) \newline c = 13.904(9) & Bi \newline Fe \newline O & 6a \newline 6a \newline 18b & 0.0 \newline 0.0 \newline 0.4486(5) & 0.0 \newline 0.0 \newline 0.0208(5) & 0.0 \newline 0.2212(3) \newline 0.9430(2) & Bi-O \newline Bi-O \newline Fe-O \newline Fe-O & 2.114(3) \newline 2.673(2) \newline 1.877(3) \newline 2.234(4) & Fe-O-Fe \newline O-Bi-O & 150.95(5) \newline 78.40(4) & 27.2 & 20.4 & 16.1 \\  
\\
\hline \hline 

\end{tabular}
\end{table*}

\end{document}